\documentclass[english,aps,pra,twocolumn,showpacs,superscriptaddress]{revtex4}
\usepackage[T1]{fontenc}
\usepackage[latin9]{inputenc}
\usepackage{amsmath}
\usepackage{amssymb}
\usepackage{graphicx}
\usepackage{babel}

\begin{document}

\title{Indirect control of spin precession by electric field via spin-orbit
coupling}

\author{Li-Ping Yang}

\affiliation{State Key Laboratory of Theoretical Physics, Institute of Theoretical
Physics and University of the Chinese Academy of Sciences, Beijing
100190, People's Republic of China}
\affiliation{Synergetic Innovation Center of Quantum Information and Quantum Physics,
University of Science and Technology of China, Hefei, Anhui 230026, China}

\author{C. P. Sun}

\email{cpsun@csrc.ac.cn}

\homepage{http://www.csrc.ac.cn/~suncp/}

\affiliation{Beijing Computational Science Research Center, Beijing 100084, China}
\affiliation{Synergetic Innovation Center of Quantum Information and Quantum Physics,
University of Science and Technology of China, Hefei, Anhui 230026, China}

\date{\today}

\begin{abstract}
The spin-orbit coupling (SOC) can mediate electric-dipole spin resonance
(EDSR) with an a.c. electric field. By applying a quantum linear coordinate transformation,
we find that the essence of EDSR could be understood as an spin precession under an effective a.c. magnetic
field induced by the SOC in the reference frame, which is exactly
following the classical trajectory of this spin. Based on this observation,
we find a upper limit for the spin-flipping speed in the EDSR-based control of spin.
For two-dimensional case, the azimuthal dependence of the effective magnetic field can be used
to measure the ratio of the Rashba and Dresselhaus SOC strengths.
\end{abstract}


\maketitle

\section{introduction}

It is of great importance to prepare and manipulate the pure quantum
state of single particle for quantum information processing and even
for the future quantum devices using the new degrees of freedom, such
as spin. Through Coulomb blockade, the single electron state in the
charge degree of freedom has been realized in quantum dot (QD) system~\cite{single-spin,single-spin_4,single-spin_5,single-spin_6,single-spin_7,single-spin_8,single-spin_9,single-spin_10}.
In the past decade, the spintronics provides a new paradigm for quantum
operations of spin in addition to the electric charge~\cite{spintronics_1,spintronics}.
However, how to perfectly control the quantum state of a single spin
in a QD is still challenging.

The conventional technology to flip spin is based on the electron
spin resonance (ESR)~\cite{ESR}, whereby the resonant magnetic field
pulses are applied. Different from a.c. electric field, which could
be generated by exciting a local gate electrode, the strong and high-frequency
magnetic field is very difficult to be applied to a micro/nano-structure
with QD effectively~\cite{ESR_QD,ESR_QD_1}. To overcome this problem,
physicists try to control electron spin in another fashion with electric
field. Rashba and Efros~\cite{EDSR_T,EDSR_T0} proposed to realize
indirect control of the electron spin by electric field through the
spin-orbit coupling (SOC)~\cite{Dressellhause,Rashba}. With SOC,
the moving electron spin seems to experience an additional effective
magnetic field induced by the a.c. electric field. When the frequency
of the electric field matches the Zeeman splitting of the electron
spin, the coherent control of a single electron spin can be achieved
with electric field indirectly~\cite{EDSR_E2D,EDSR_Eprl07,EDSR_Enp,EDSR_E1D,EDSR_E2D13}.
This spin resonance in the effective magnetic field is called electric-dipole
spin resonance (EDSR)~\cite{EDSR}.

There are plenty of literatures~\cite{EDSR_T-1,EDSR_T-2,EDSR_T-3,EDSR_T-4,EDSR_T-5}
on the theoretical explanation of EDSR effect. An intuitive picture
of EDSR is given by V. N. Golovach\textit{ et al.}~\cite{EDSR_T-1}.
They first eliminated the SOC terms by making the Schrieffer-Wolff
transformation to the first order of perturbation theory, and then they found
that the electric field would behave as an effective magnetic field.
In this paper, we revisit this enlightening physical explanation by
studying the spin dynamics in a reference frame, which exactly follows
the classical trajectory of a driven electron trapped in a harmonic
potential. For the QD in a one-dimensional (1D) nanowire, the electron
is constrained in a 1D harmonic trap and driven by a.c. electric field.
When the trap is tight enough, the influence of the high-frequency
free oscillation of the electron to the dynamics of the spin can be
neglected in the reference frame co-moving with the electron, but the forced
oscillation of the electron under the a.c. electric field with lower frequency
provides the spin with an effective resonant magnetic field through SOC.
If the direction of the electric field is along the wire, the magnetic field induced
by Rashba SOC is perpendicular to the electric field, while the one
induced by Dresselhaus SOC is parallel to the electric field. The
induced magnetic field has the same frequency as the driving electric
field, and then the coherent controlling of the spin can be realized
when the electric field is resonant with the Zeeman splitting ($\omega_z$) of the
electron spin. Our investigation here shows that for a tight trap,
one can enlarge the Zeeman splitting of the electron spin, 
in addition to increasing SOC or electric driving strength, 
to increase the spin-flip speed. But there exists an upper
limit of the effective Rabi frequency $(10^{-3}-10^{-2})\omega_z$
of the coherent control of the spin with EDSR~\cite{PRB12}.

For the two-dimensional (2D) QD system, we find that the induced a.c. magnetic
field becomes azimuth dependent. If the external static magnetic field is weak and
the 2D harmonic well is isotropic, we discover that the a.c. magnetic
field induced by Dresselhaus SOC and the a.c. electric field lie
at two different sides of the $x$-axis with the same angle from the
$x$-axis and the magnetic field induced by Rashba SOC is still perpendicular
to the electric field. Based on the above understanding about the EDSR, we can realize the
precise control of spin precession on the Bloch sphere surface. On the other hand,
we can also measure the ratio of the Rashba and Dresselhaus SOC strengths by using of this
azimuthal dependence of the induced magnetic field as proposed in Ref~\cite{EDSR_T0}.

\begin{figure}
\centering
\includegraphics[width=8.5cm]{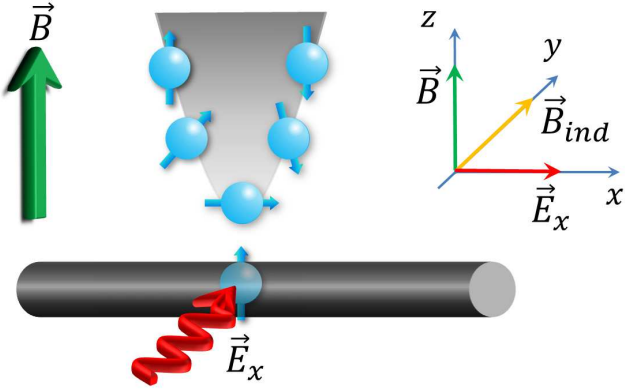}
\caption{(Color online) The electron moving in a one-dimensional nanowire is
constrained in a harmonic trap. The electron spin is initially polarized
along the $z$-axis by an external static magnetic field $\vec{B}$.
When an a.c. electric field $\vec{E}_{x}$ is applied to drive the
electron, the electron spin will experience an additional magnetic
field induced by the SOC. In our case, the magnetic field induced
by Rashba SOC is perpendicular to the electric field.}
\label{fig:schematic}
\end{figure}

In the next section, we present our model and show the origin of the effective magnetic field in EDSR.
In Sec.~III, the effective spin precession under the electric field through SOC is presented.
We investigate how to speed up the coherent spin control with EDSR in Sec.~IV.
In Sec.~V, we study the coherent spin control via EDSR in 2D QD system.
Finally, the summary of our main results is given in Sec.~VI.
Some detailed calculations are displayed in the Appendices.

\section{the electric-dipole spin resonance}

We first take the 1D
nanowire QD system, where an electron is confined in an 1D harmonic
trap along the $x$-direction with frequency $\omega$ (Fig.~\ref{fig:schematic}), as an illustration
to explore the physical mechanism of EDSR. The model Hamiltonian reads~\cite{Hamiltonian_1,Hamiltonian_2},
\begin{equation}
H=\hbar\omega a^{\dagger}a-i\tilde{\alpha}(a^{\dagger}-a)\sigma_{y}+\frac{1}{2}\hbar\omega_{z}\sigma_{z}+\xi(a^{\dagger}+a)\cos\nu t,\label{eq:H_original}
\end{equation}
where $a=\sqrt{m_{e}\omega/(2\hbar)}[x+ip_{x}/(m_{e}\omega)]$ is
the annihilation operator of the vibration degree of the electron
with coordinate (momentum) $x\ (p_{x})$ and effective mass $m_{e}$,
$\tilde{\alpha}=\alpha\sqrt{m_{e}\omega/(2\hbar)}$ with the Rashba
SOC constant $\alpha$, $\xi=eE_{x}\sqrt{\hbar/(2m_{e}\omega)}$ is
the effective driven strength of the a.c. electric field $-E_{x}\cos\nu t$.
An external static magnetic field with strength $B$ is applied along
the $z$-direction to to polarized the electronic spin. Here, $\omega_{z}=g\mu_{B}B/\hbar$
is the Zeeman splitting of the electron spin with the Bohr magneton
$\mu_{B}$ and the effective $g$-factor $g$. Here, we just take
the Rashba SOC into account, because our approach can be generalized
to Dresselhaus case straightforwardly. Hereafter, we take $\hbar=1$
for convenience.

Now, we consider the dynamics of spin precession in a reference frame
exactly following the classical trajectory of the driven electron
in a harmonic trap. To this end, we introduce the time-dependent displacement
transformation~\cite{Cadez}
\begin{equation}
D[f(t)]=e^{f(t)a^{\dagger}-{\rm h.c.}}\equiv e^{-i[p_{x}x_{c}(t)+xp_{c}(t)]},
\end{equation}
where $p_{c}(t)=m\dot{x}_{c}$ and
\[
x_{c}(t)=-\sqrt{\frac{2}{m_{e}\omega}}\frac{\omega\xi}{\omega^{2}-\nu^{2}}\cos\nu t,
\]
corresponds to the classical trajectory of a driven harmonic oscillator
(DHO) described by the classical Hamiltonian
$\mathcal{H}_{c}=p_{c}^{2}/(2m_{e})+m\omega^{2}x_{c}^{2}/2+\sqrt{2m_{e}\omega}\xi x_{c}\cos\nu t$
(please refer to Appendix A for details).
Here, $f(t)=-(\sqrt{m_{e}\omega/2}x_{c}+ip_{c}/\sqrt{2m_{e}\omega})$
represents a complex displacement in the phase space. As displayed in Appendix A,
the above unitary transformation
$D(t)=D[f(t)]$ is equivalent to the quantum linear coordinate transformation~\cite{Leblond}
\begin{equation}
x'=x-x_{c}(t),\ t'=t,\ \nabla'=\nabla,\ \frac{\partial}{\partial t'}=\frac{\partial}{\partial t}+\dot{x}_{c}\nabla',
\end{equation}
accompanied by a corresponding transformation of the wave function
$\psi'(x',t')=\psi(x,t)\rm{exp}(-i\phi)$,
with $\phi=[p_{c}x'+(1/m_e)\int_{0}^{t'}p_{c}^{2}(\tau)d\tau]$,
as the requirement of covariance. The above transformation $D[f(t)]$
gives the equivalent Hamiltonian $H_{D}=DHD^{\dagger}-iD\left(\partial_{t}D^{\dagger}\right)$
in the reference frame moving along the classical path $x_{c}(t)$
as
\begin{eqnarray}
H_{D} & \!\!=\!\! & \omega a^{\dagger}a\!-\! i\tilde{\alpha}(a^{\dagger}\!-\! a)\sigma_{y}\!-\!\eta\tilde{\alpha}\sigma_{y}\sin\nu t\!+\!\frac{1}{2}\omega_{z}\sigma_{z}.\label{eq:H_D}
\end{eqnarray}
Here, $\eta=2\nu\xi/(\omega^{2}-\nu^{2})$
is a dimensionless parameter and we have neglected a time-dependent
c-number $\mathcal{E}_{c}=-(\omega\xi^{2}\cos^{2}\nu t)/(\omega^{2}-\nu^{2})$,
which corresponds to the classical energy of this DHO.

It is found that $\eta$ is proportional to the driving strength $\xi$
and will be greatly enhanced if the driving frequency $\nu$ is nearly
resonant with the frequency of the trap $\omega$. It follows Hamiltonian~(\ref{eq:H_D})
that, in the new reference frame, a time-dependent spin flipping term appears,
and its frequency is the same as the a.c. electric field.

\section{effective spin precession.}

In experiment~\cite{EDSR_E1D},
the electron is tightly constrained in the trap with orbital transition
energy $\sim5-9{\rm meV}$ (corresponding oscillating frequency $\omega\sim10^{12}-10^{13}{\rm Hz}$),
the SOC strength $\tilde{\alpha}\sim10^{9}{\rm Hz}$, the Zeeman splitting
$\omega_{z}$ is of $\sim10^{10}{\rm Hz}$ with $B\sim100{\rm mT}$
and $g\approx9$, and the driving frequency $\nu$ is resonant with
$\omega_{z}$. Thus we have the condition $\omega\gg\omega_{z}\approx\nu>\tilde{\alpha}$,
and then we can adiabatically eliminate the degree of the freedom
of the vibration part to obtain the effective spin Hamiltonian.

The formal solution of the Heisenberg equation of $a(t)$ reads
\begin{equation}
a(t)=a(0)e^{-i\omega t}-\tilde{\alpha}e^{-i\omega t}\int_{0}^{t}\sigma_{y}(\tau)e^{i\omega\tau}d\tau.\label{eq:solution_a}
\end{equation}

In the case of $\omega\gg\omega_{z}\approx\nu>\tilde{\alpha}$, the
oscillating frequency of the spin operator is of the scale
$\sim \exp(\pm i\omega_z t)$, then the magnitude of the
integral in Eq.~(\ref{eq:solution_a}) is approximated as~$\tilde{\alpha}/(\omega\pm\omega_z)\ll 1$.
As a result, the influence of the SOC to the dynamics of the vibration
part can be neglected. Then we can take the semi-classical approximation
by replacing $a$ and $a^{\dagger}$ in Hamiltonian (\ref{eq:H_D})
with $\langle a(0)\rangle_{D}\exp(-i\omega t)$ and $\langle a^{\dagger}(0)\rangle_{D}\exp(i\omega t)$
($\langle\cdots\rangle_{D}$ means averaging over the displaced initial
state), respectively. For simplicity, we assume the system is in the
state $\left|\psi(0)\right\rangle =\left|0\right\rangle \otimes\left|\uparrow\right\rangle $,
and then the effective Hamiltonian for the spin part reads
\begin{equation}
H_{s}^{{\rm eff}}=\frac{1}{2}\omega_{z}\sigma_{z}-\eta\tilde{\alpha}\sigma_{y}\sin\nu t+\frac{2\omega}{\nu}\eta\tilde{\alpha}\sigma_{y}\sin\omega t.\label{eq:H_eff_s}
\end{equation}

Then the influences of the forced oscillation under the electric field
and the free oscillation of the electron to the spin are described
by two effective a.c. magnetic fields $B_{{\rm E}}=-B_{{\rm ind}}\sin\nu t$
and $B_{{\rm free}}=(2\omega/\nu)B_{{\rm ind}}\sin\omega t$,
with $B_{{\rm ind}}=\eta\tilde{\alpha}/(g\mu_{B})$.
The strengths of these two effective magnetic fields are both proportional
to $\tilde{\alpha}\xi$. In 1D case, the effective magnetic fields
induced by Rashba SOC (both $B_{{\rm free}}$ and $B_{{\rm E}}$)
are perpendicular to the electric field, and the magnetic fields induced
by the Dresselhaus SOC are parallel to the electric field.

We find that the effective magnetic field $B_{{\rm E}}$ generated
by the forced oscillation of the electron under the electric field
through SOC has the same frequency $\nu$ as the electric field. The
effective field $B_{{\rm free}}$ generated by the free oscillation
of the electron has the same frequency as the frequency of the harmonic
trap $\omega$. Although the strength of $B_{{\rm free}}$ is $2\omega/\nu$
times lager than $B_{{\rm E}}$, it contribute little to the spin
control, because its frequency $\omega$ is largely detuning from
$\omega_{z}$. Therefore, in the case of $\omega\gg\omega_{z}\approx\nu>\tilde{\alpha}$,
the dynamics of the spin can be described by the following Hamiltonian,
\begin{equation}
H_{s}^{{\rm RWA}}=\frac{1}{2}\omega_{z}\sigma_{z}-\frac{\eta\tilde{\alpha}}{2}(\sigma_{+}e^{-i\nu t}+\sigma_{-}e^{i\nu t}).\label{eq:H_RWA}
\end{equation}
Here, we have neglected the influence of the fast oscillating term
$B_{{\rm free}}$ and taken the rotating approximation (RWA) of the
resonant term associated with $B_{{\rm E}}$.

\begin{figure}
\centering

\includegraphics[width=6cm]{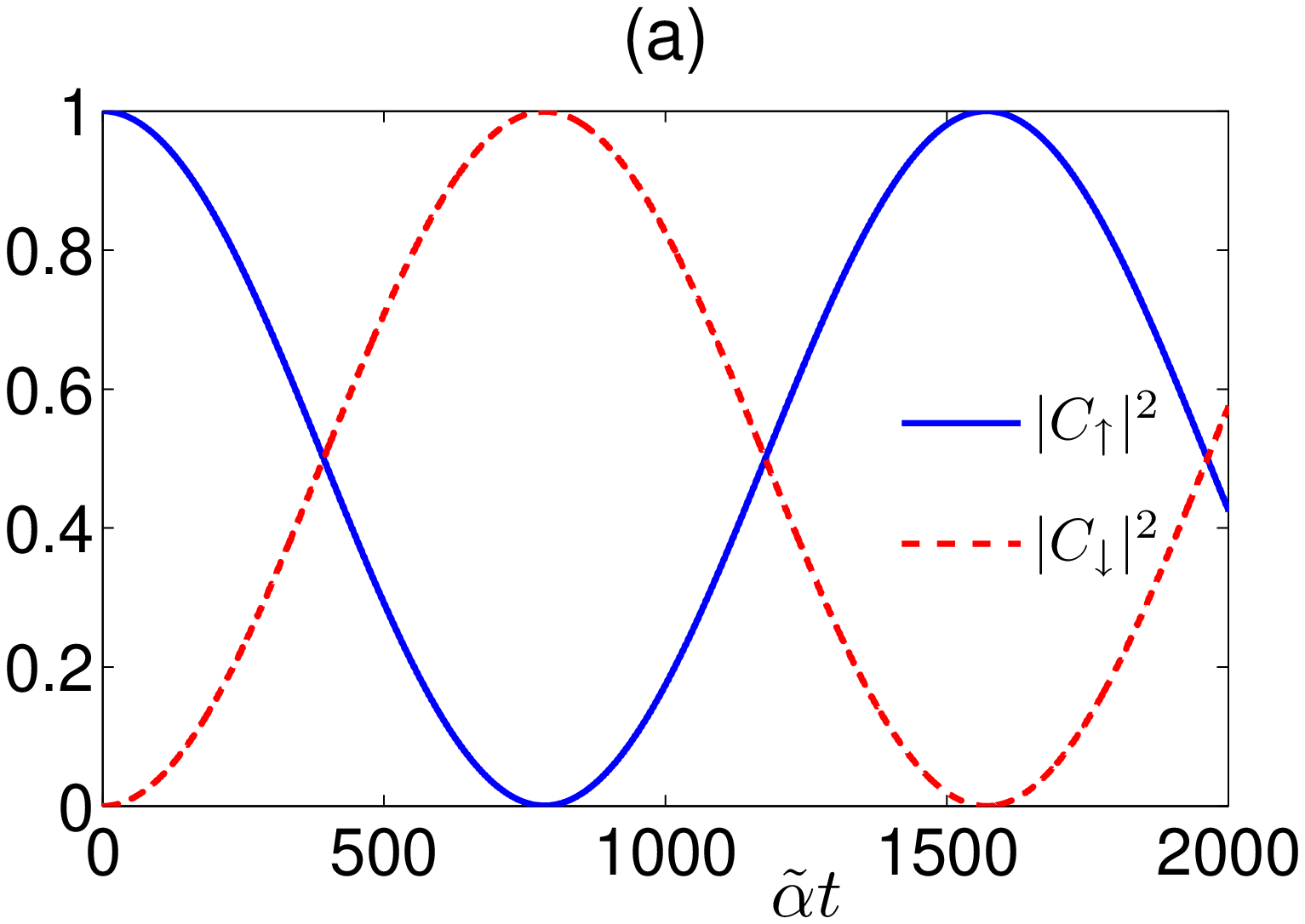}

\includegraphics[width=6cm]{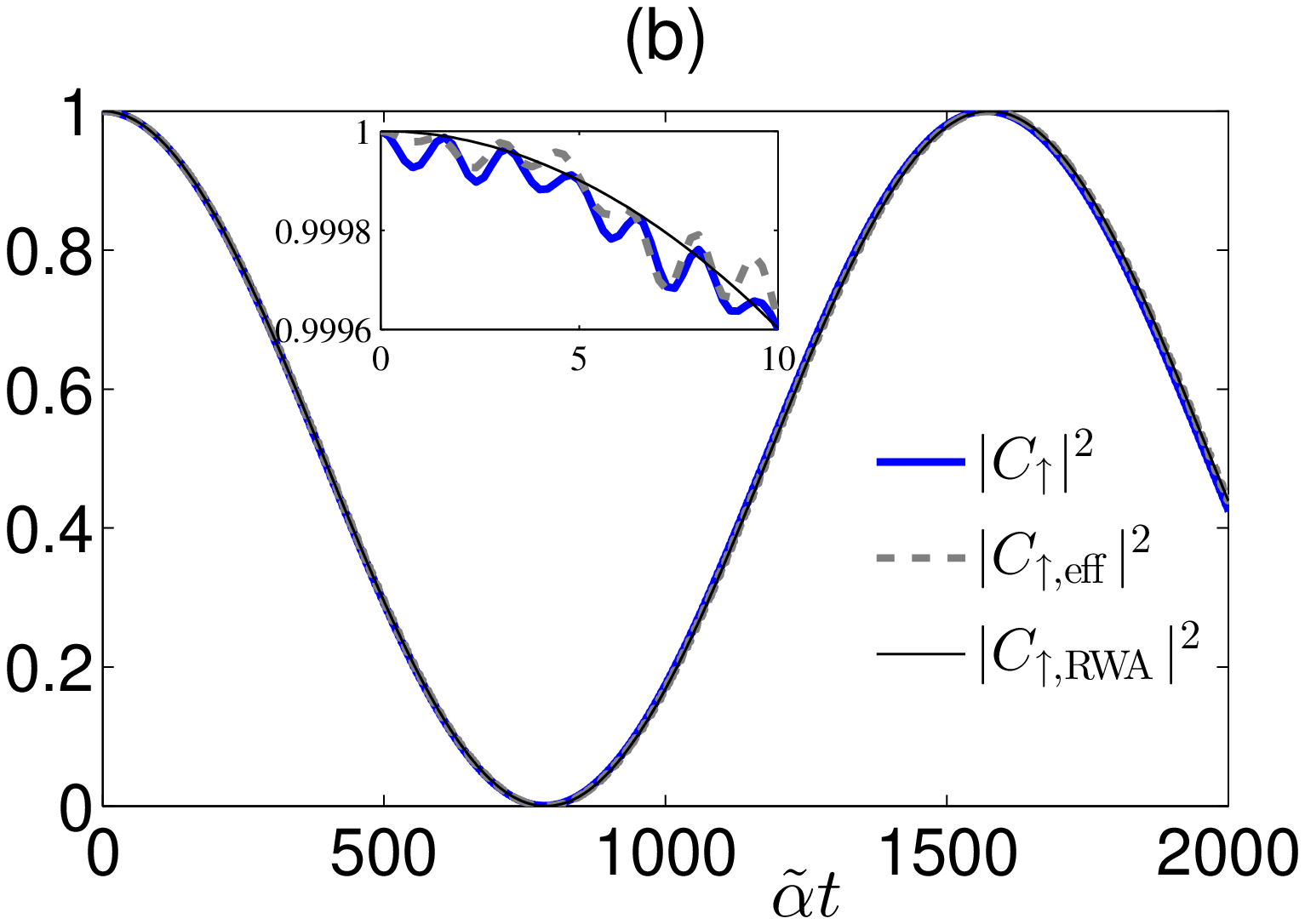}

\caption{(Color online) We take the SOC constant as the unit $\tilde{\alpha}=1$.
$\omega=250$, $\omega_{z}=2.5$, $\nu=2.5$, and $\xi=50$. (a) The
solid blue (dashed red) line is the probability of the $\left|\uparrow\right\rangle $
($\left|\downarrow\right\rangle $) state of the spin, which is directly
calculated via $H_{D}$. (b) Probabilities of the spin state $\left|\uparrow\right\rangle $
obtained by different methods are presented. Solid blue line is obtained the numerical calculation
via $H_{D}$. The dashed gray line is obtained from $H_{s}^{{\rm eff}}$.
And the thin black line is obtained from $H_{s}^{{\rm RWA}}$ with
rotating wave approximation.}
\label{fig:fig2}
\end{figure}

To demonstrate the coherent control of the electron spin, we turn
to numerical calculations. The spin is initially polarized in the
state $\left|\uparrow\right\rangle $ and its wave function at time
$t$ can be expanded as $\left|\chi(t)\right\rangle =C_{\uparrow}(t)\left|\uparrow\right\rangle +C_{\downarrow}(t)\left|\downarrow\right\rangle $.
Here, $|C_{\uparrow}(t)|^{2}$ ($|C_{\downarrow}(t)|^{2}$) denotes
the occupation probability of the state $\left|\uparrow\right\rangle $
($\left|\downarrow\right\rangle $). In the case of $\omega\gg\omega_{z}=\nu>\tilde{\alpha}$,
the perfect Rabi oscillation of the spin with frequency $\eta\tilde{\alpha}\sim(10^{-4}-10^{-3})\tilde{\alpha}$
is observed when the driving frequency $\nu$ is resonant with $\omega_{z}$
as shown in Fig.~\ref{fig:fig2} (a). In Fig.~\ref{fig:fig2}~(b), the solid blue line is obtained
directly from the Hamiltonian $H_{M}$ by tracing off the degree of
freedom of the vibration, the dashed gray line is obtained from $H_{s}^{{\rm eff}}$,
and thin black line is from $H_{s}^{{\rm RWA}}$.
These three lines coincide with each other very well, except for some
high-frequency fluctuation around the thin black line (obtained from
$H_{s}^{{\rm RWA}}$) as shown in the subgraph of Fig.~2 (b). Thus,
the dynamics of the spin is well described by $H_{s}^{{\rm RWA}}$
in the regime $\omega\gg\omega_{z}=\nu>\tilde{\alpha}$. And then
the electron spin could be well controlled with electric field via
EDSR as the same as ESR.

\section{coherent-spin-control speed enhancement}

In the preceding sections, we find that one can coherent control the electric spin
with electric field through EDSR just like with magnetic field. Next, we will explore
how to speed up coherent spin control with EDSR. First, we gives the condition to realize 
coherent control of electron spin with EDSR. For a spin-$1/2$
system described by Hamiltonian (\ref{eq:H_RWA}), the flipping probability
$|C_{\downarrow}(t)|^{2}$ for the Rabi oscillation reads
\begin{equation}
|C_{\downarrow}(t)|^{2}=\frac{(\eta\tilde{\alpha})^{2}}{\delta^{2}+(\eta\tilde{\alpha})^{2}}\sin^{2}\left(\sqrt{\delta^{2}+(\eta\tilde{\alpha})^{2}}t/2\right),\label{eq:Rabi}
\end{equation}
where $\delta=\omega_{z}-\nu$ is the detuning between the driven
frequency and the Zeeman splitting of the electron spin. As shown in Fig.~3 (a), the amplitude
of the Rabi oscillation tends to $1$ when $\delta\ll\eta\tilde{\alpha}$,
while the amplitude tends to $0$ when $\delta\gg\eta\tilde{\alpha}$.
As a result, the flipping probability is deeply suppressed by the detuning
$\delta$.

\begin{figure}
\centering
\includegraphics[width=6cm]{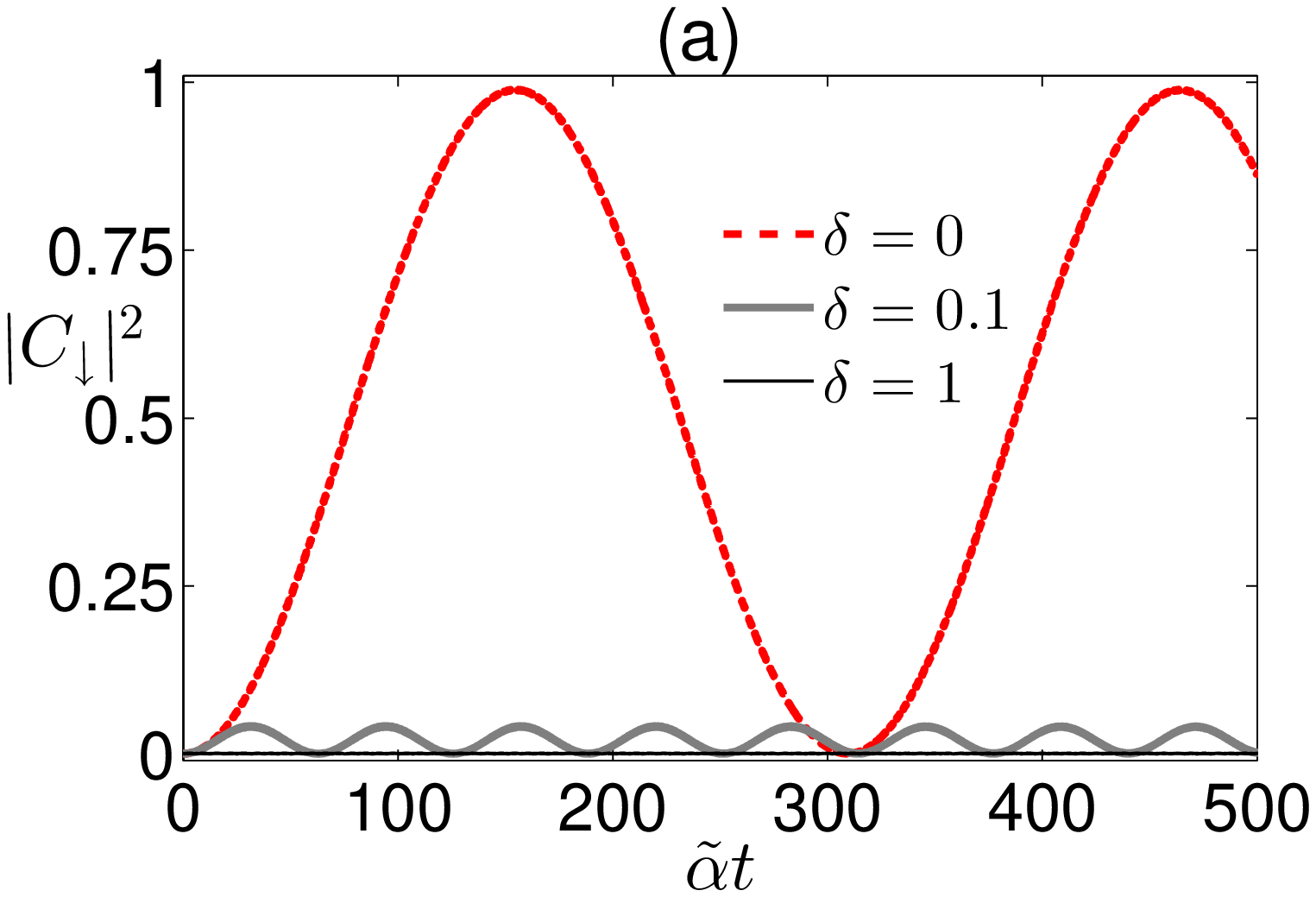}
\includegraphics[width=6cm]{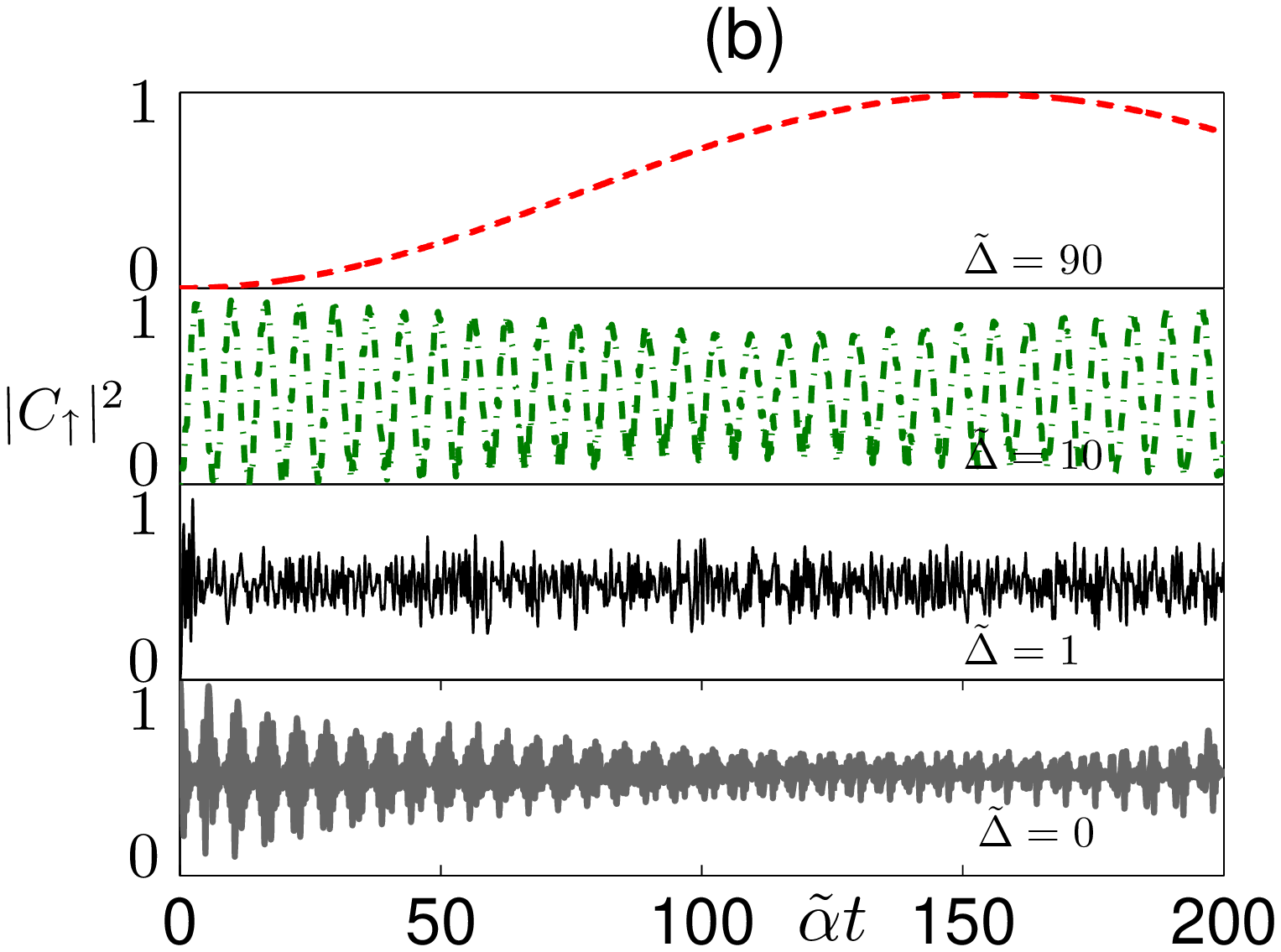}
\caption{(Color online) Here, we take the SOC strength as the unit $\alpha=1$.
(a) The flipping probabilities for different detunings
$\delta$ between the driving electric field and the Zeeman splitting
of the electron spin are presented. The parameters are taken as $\omega=100$, $\omega_z=10$,
$\xi=10$. The dotted red line described the resonant
case with $\nu=\omega_{z}$, $\delta=0$ and $\eta=0.0202$. The solid
gray line described the case with $\nu=9.9$, $\delta=0.1$ and $\eta=0.02$.
The thin black line described the large detuning case with $\nu=9$,
$\delta=1$ and $\eta=0.0181$. (b) The flipping probabilities for different detunings $\tilde{\Delta}$
between frequency of the harmonic well and the Zeeman splitting of
the electron spin are presented. Here, $\omega=100$, $\xi=10$, and
the resonance condition $\delta=0$ is always guaranteed. And the other
parameters are: $\nu=\omega_{z}=10$ with $\tilde{\Delta}=90$ for
dotted red line, $\nu=\omega_{z}=90$ with $\tilde{\Delta}=10$ for
dash-dotted green line, $\nu=\omega_{z}=99$ with $\tilde{\Delta}=1$
for thin black line, $\nu=\omega_{z}=100$ with $\tilde{\Delta}=0$
for solid gray line.}
\label{fig:fig3}
\end{figure}

Similarly, when the detuning between the frequencies of harmonic trap
and the Zeeman splitting $\tilde{\Delta}=\omega-\omega_{z}\gg\omega\eta\tilde{\alpha}/\nu$
is large, the effective magnetic field $B_{{\rm free}}$ hardly affect
the spin-flipping process. As a result, the influence of the free
oscillation of the electron to the dynamics of the spin can be neglected
in the former section.

Starting from the original Hamiltonian (\ref{eq:H_original}), we numerically
study the the influence of the detuning $\tilde{\Delta}$ to the coherent
controlling of the spin. It is discovered that when $\tilde{\Delta}\gg\tilde{\alpha}$
and $\delta\ll\eta\tilde{\alpha}$, the coherent controlling of electron
spin is realized as shown by the dotted red line ($\tilde{\Delta}=90 \tilde{\alpha}$)
and dashed-dot green line ($\tilde{\Delta}=10 \tilde{\alpha}$) in Fig.~\ref{fig:fig3}~(b). When
$\tilde{\Delta}\lesssim\tilde{\alpha}$, the SOC will destroy the
coherence of the spin, and then collapse and revival phenomenon appear.
As shown by the thin black line ($\tilde{\Delta}=1\tilde{\alpha}$) and solid gray
line $\tilde{\Delta}=0$ in Fig.~4, the coherent controlling of the
electron spin is destroyed by SOC. As a result, to realize a perfect spin control
through EDSR, two necessary conditions must be guaranteed:
(1) the frequency of the a.c. electric field must be resonant with
the Zeeman splitting of the spin in the external magnetic field, i.e., $\delta\ll \eta \tilde{\alpha}$;
(2) the frequency of the harmonic trap must be largely detuned from
the Zeeman splitting of the spin, i.e.,~$|\tilde{\alpha}/\tilde{\Delta}|\ll 1$.

\begin{figure}
\centering
\includegraphics[width=6cm]{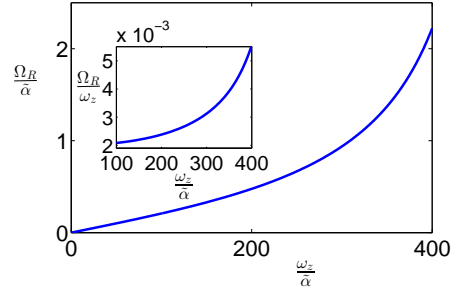}
\caption{(Color online) Here, we take the SOC strength as the unit $\alpha=1$.
The Rabi frequency of the coherence spin flip increases with $\omega_z$.
The parameters are taken as $\xi=\omega=500$.}
\label{fig:fig4}
\end{figure}

According to Eq.~(\ref{eq:Rabi}), spin-flip speed is charaterized by the Rabi frequency:
\begin{equation}
\Omega_{R}=\frac{1}{2}\eta\tilde{\alpha}=\frac{\omega_z \xi \tilde{\alpha}}{\omega^2-\omega_{z}^{2}},\label{eq:OmegaR}
\end{equation}
where we have used the resonance condition $\delta=\omega_z-\nu=0$. It is found that this Rabi frequency~$\Omega_R$ is proportional to the strengths of the SOC $\tilde{\alpha}$ and electric
driving $\xi$. Hence, we should increase $\tilde{\alpha}$ and $\xi$ to enlarge~$\Omega_R$.
But if the driving strength~$\xi\gg\omega$, the electron will flee from the harmonic
trap. When~$\tilde{\alpha}$ is large enough to break the large detuning
condition~$|\tilde{\alpha}/\tilde{\Delta}|\ll 1$, the free oscillation of the electron
and the spin precession will be highly
correlated, thus the SOC will destroy the coherent control according
to Hamiltonian~(\ref{eq:H_D}). For safety, we require~$\xi\leqslant\omega$
and~$0<\tilde{\alpha}/\tilde{\Delta}\leqslant 0.01$ to guarantee the perfect Rabi oscillation of
the electron spin in the trap. From Eq.~(\ref{eq:OmegaR}), $\Omega_R$ monotonically increases with $\omega_z$
with  upper limit $10^{-2} \omega_z$ for this case (see Fig.~\ref{fig:fig4}).
Thus, for a very large~$\omega$, one can increase~$\omega_z$, besides the driving strength~$\xi$
and SOC strength~$\tilde{\alpha}$, to speed up the spin flipping.

\section{two-dimensional Quantum Dot system}

For the two-dimensional (2D)
QD system, the magnetic field induced by the a.c. electric field
via SOC becomes much more complicated and azimuth-dependent. 
We can utilize this azimuth dependence to measure the Rashba and
Dresselhaus SOC strength ratio and realize a perfect single electron
spin qubit operation through EDSR.

The Hamiltonian of the electron confined in an 2D harmonic well
$H=H_{v}+H_{s}+H_{{\rm so}}+V(t)$ is composed of four parts: the
vibration part of the electron is described with
\begin{equation}
H_{v}=\frac{1}{2m_{e}}[\vec{p}+\frac{e}{c}\vec{A}(\vec{r})]^{2}+\frac{1}{2}m_{e}\tilde{\omega}_{x}^{2}x^{2}+\frac{1}{2}m_{e}\tilde{\omega}_{y}^{2}y^{2},\label{Hv}
\end{equation}
where $\vec{p}=p_{x}\vec{e}_{x}+p_{y}\vec{e}_{y}$ is the in-plane
momentum, $\tilde{\omega}_{x(y)}$ the frequency of the harmonic trap
of $x(y)$-direction, and $\vec{A}(\vec{r})=B(0,0,y\cos\varphi_{B}-x\sin\varphi_{B})$
the vector potential for the in-plane static magnetic field $\vec{B}=B(\cos\varphi_{B},\sin\varphi_{B},0)$.
The second part~$H_{s}=\hbar g\mu_{B}\vec{B}\cdot\vec{\sigma}/2$ describes the Zeeman splitting of the electron
spin in $\vec{B}=B(\cos\varphi_{B},\sin\varphi_{B},0)$.
For the third part, both the Rashba and Dresselhaus SOC with strength $\alpha_{R}$
and $\alpha_{D}$ respectively are taken into account
\begin{equation}
H_{{\rm so}}=\alpha_{R}(\sigma_{x}p_{y}-\sigma_{y}p_{x})+\alpha_{D}(\sigma_{y}p_{y}-\sigma_{x}p_{x}).
\end{equation}
The last part~$V(t)=-e\vec{r}\cdot\vec{E}(t)$ describes the driven of the electron under an
in-plane a.c. electric field $\vec{E}(t)=-\left(E\vec{e}_{x}\cos\varphi_{E}+E\vec{e}_{y}\sin\varphi_{E}\right)\cos\nu t$.

\begin{figure}
\centering
\includegraphics[height=5cm]{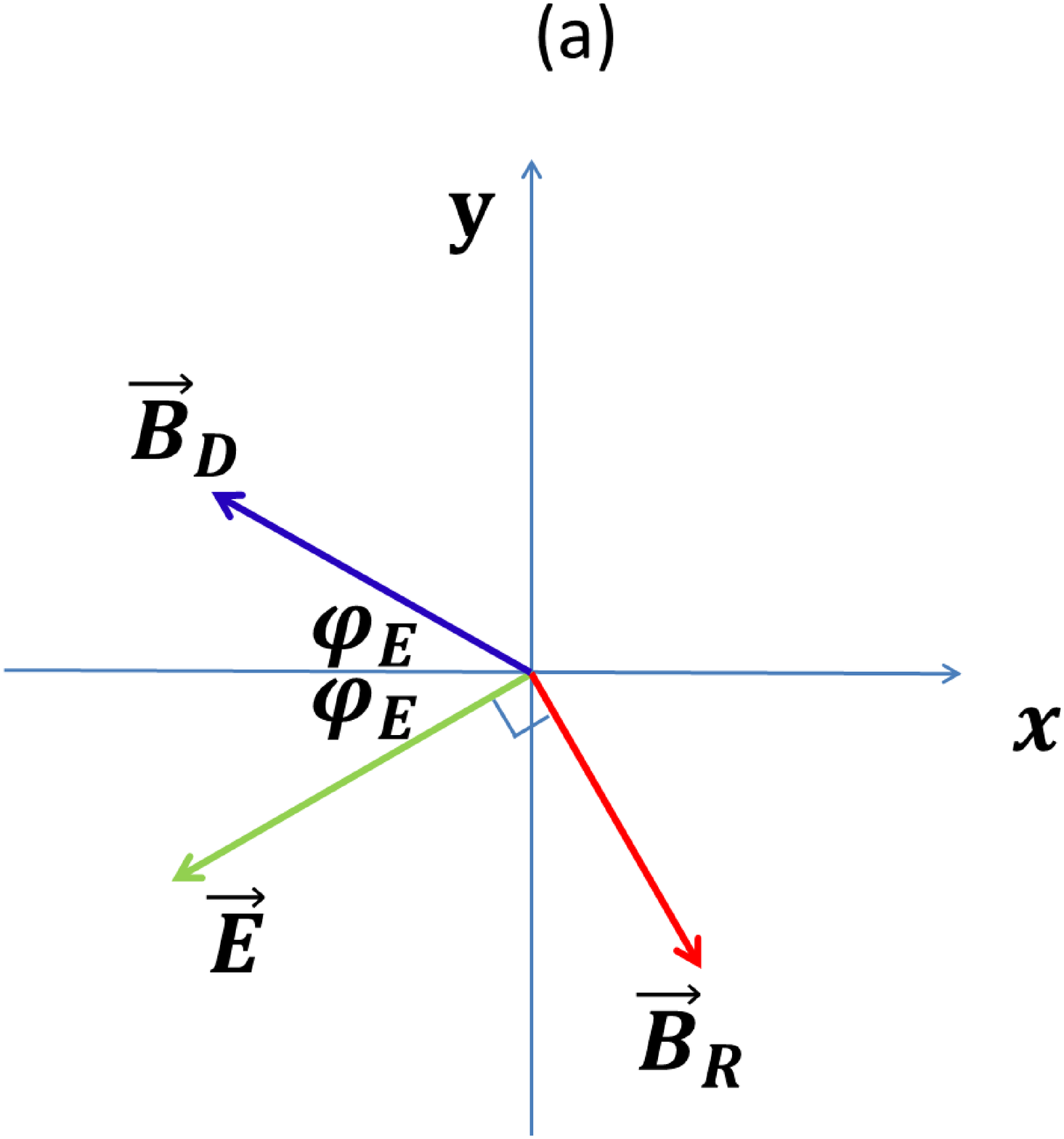}\includegraphics[height=5cm]{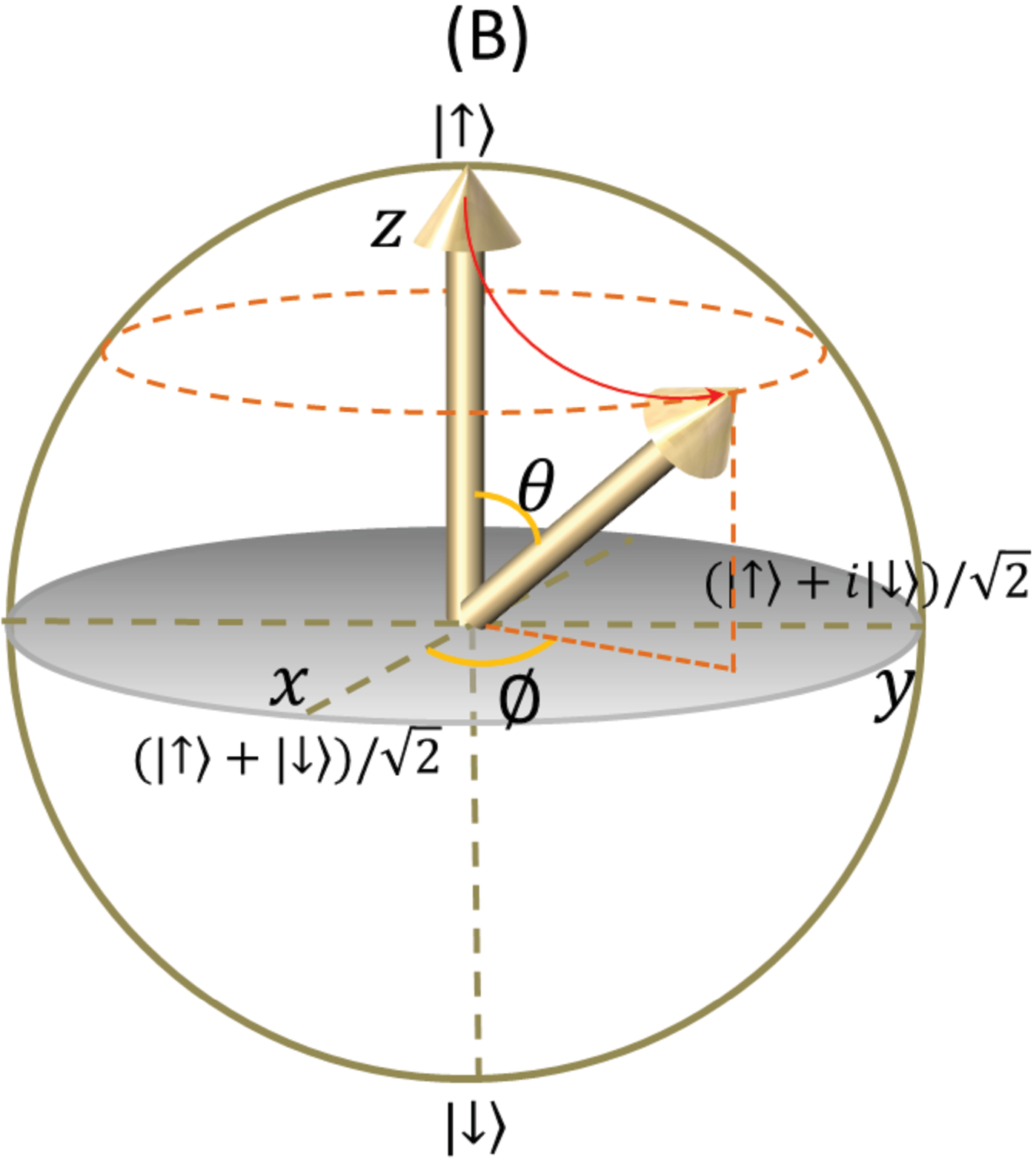}

\caption{(Color online) The 2D electric field caused spin precession on the
Bloch sphere. (a) The a.c. magnetic field induced by Dresselhaus
SOC and the a.c. electric field lie at two different sides of the
$x$-axis with the same angle from the $x$-axis. The magnetic field
induced by Rashba SOC $\vec{B}_{R}$ is always vertical to $\vec{E}$.
(b) The trajectory  of the spin from the initial point
$\rho(0)=\left|\uparrow\right\rangle \left\langle \uparrow\right|$
to the final state $\rho(\theta,\phi)$ with $\theta=g\mu_{B}B_{R}t/2$
and $\phi=\nu t+\varphi_{E}$ on the Bloch sphere surface.}
\label{fig:fig5}
\end{figure}

After a similar unitary transformation as the 1D case, the a.c. electric
field will be converted to an a.c. magnetic field with the same frequency
(please refer to Appendix B for details). In
the experiment~\cite{EDSR_E2D}, the external static magnetic field
is week, i.e., the frequency modification induced by the vector potential
can be neglected $\omega_{c}=eB/(m_{e}c)\ll\tilde{\omega}_{x(y)}$~\cite{NZhao}. In this
case, the effective spin Hamiltonian for the isotropic 2D harmonic
well $\tilde{\omega}_{x}=\tilde{\omega}_{y}=\omega$ could be simplified
to
\begin{equation}
H_{s}^{{\rm eff}}=\frac{1}{2}g\mu_{B}\left[\vec{B}+\vec{B}_{R}(t)+\vec{B}_{D}(t)\right]\cdot\vec{\sigma},
\end{equation}
where $\vec{B}_{R}(t)=B_{R}\sin\nu t(\sin\varphi_{E},-\cos\varphi_{E},0)$
and $\vec{B}_{D}(t)=B_{D}\sin\nu t(-\cos\varphi_{E},\sin\varphi_{E},0)$
are the magnetic fields induced by Rashba and Dresselhaus SOC respectively,
with strength
\begin{equation}
B_{R(D)}(t)=\frac{2\hbar\nu eE\alpha_{R(D)}}{g\mu_{B}\left(\omega^{2}-\nu^{2}\right)}.
\end{equation}
It is observable that $\vec{B}_{R}$ is always vertical to the electric
field $\vec{E}$, but $\vec{B}_{D}$ is vertical to $\vec{E}$ only
when $\varphi_{E}=n\pi/2+\pi/4\ (n=0,1,2,3)$ [see Fig.~\ref{fig:fig5} (a)].
When $\varphi_{E}=3\pi/4\ {\rm or}\ 7\pi/4$,
$\vec{B}_{R}$ and $\vec{B}_{D}$ are parallel, while $\vec{B}_{R}$
and $\vec{B}_{D}$ are anti-parallel when $\varphi_{E}=\pi/4\ {\rm or}\ 5\pi/4$.
Therefore, one can measure the ratio of the SOC strength $\alpha_{R}/\alpha_{D}$
by measuring the different Rabi frequencies for $\varphi_{E}=\pi/4$
and $\varphi_{E}=3\pi/4$~\cite{EDSR_T0}.

In addition, the direction of the effective magnetic field can be well controlled 
by tuning the direction fo the a.c. electronic field.
For the case where the static magnetic field along $z$-direction
$\vec{B}=B(0,0,1)$ instead of the in-plane one is applied, the induced
magnetic fields are nearly the same. If there is only one type of
SOC (e.g., Rashba one), we can use the electric field to induce an
effective magnetic field in any needed direction, which controls the evolution of the
spin state from a starting point to anywhere we wanted on the Bloch
sphere. If the electric field is resonant with the Zeeman splitting of the spin
(i.e., $\nu=g\mu_{B}B$, the dynamics of the spin can be described by the Hamiltonian
\begin{equation}
H_{s}^{RWA}\!=\!\frac{g\mu_{B}}{2}{B\sigma_{z}\!-\!\frac{B_{R}}{2}[e^{-i(\varphi_{E}+\nu t)}\sigma_{+}\!+\!e^{i(\varphi_{E}+\nu t)}\sigma_{-}]} .
\end{equation}
An arbitrary target state $\rho(\theta,\phi)=(1+\vec{n}\cdot\vec{\sigma})/2$
with $\vec{n}=(\sin\theta\cos\phi,\sin\theta\sin\phi,\cos\theta)$
on the Bloch sphere can be realized through EDSR by choosing proper
time $g\mu_{B}B_{R}t/2=\theta$ and proper angle $\varphi_{E}=\phi-\nu t$
from the starting point $\rho_{0}=\left|\uparrow\right\rangle \left\langle \uparrow\right|$
[see Fig.~\ref{fig:fig5} (b)]. Namely, the electric field can perform a perfect single-qubit
operation in the spin system through EDSR.

\section{summary}

For the physical mechanism of EDSR in 1D nanowire
QD or for 2D case, we provide an intuitive explanation with an exact
picture in physics based on the reference frame transformation: the
electric field can behaves as magnetic field in the reference frame
exactly following the classical trajectory of a DHO. This electric-magnetic duality
can be generally found in relativistic transformation of the Maxwell
equations. We also notice that SOC is in essence the consequence of
relativistic quantum theory in the low-velocity limit. Thus our approach
presented in this letter is, in principle, consistent with the point
of view of special relativity.

For the EDSR technology itself, our study shows that two necessary
conditions must be guaranteed to realize a perfect spin control
through EDSR:(1) the frequency of the a.c. electric field must be resonant with
the Zeeman splitting of the spin;
(2)the detuning beween the frequency of the harmonic trap and the Zeeman splitting
of the spin must be much larger than SOC coupling strength.
Based on these conditions, there are three
ways to increase the speed of coherent spin control: (1) increasing
the electric driving strength;(2) increasing the strength of SOC $\tilde{\alpha}$;
(3) increaseing the external state magnetic field to increase the
Zeeman splitting of the electron spin.

The azimuthal dependence of the induced magnetic field can be
used to measure the ratio of the strengths of the Rashba and Dresselhaus
SOC. We also shown that the precise control of spin in the whole Bloch
sphere can be realized in 2D QD system through EDSR technology.

We thank Da-Zhi Xu and Prof. Xia-Ji Liu for helpful discussion. This work was supported
by the National Natural Science Foundation of China Grant No.11121403 and the National 973 program (Grant No.
2012CB922104 and No. 2014CB921403).

\appendix

\section{the quantum driven harmonic oscillator }

\subsection{Time-dependent displacement transformation}

The exact solution of the Schr\"odinger equation of a quantum driven
harmonic oscillator (DHO) described by Hamiltonian
\begin{equation}
H=\omega a^{\dagger}a+[F(t)a+{\rm h.c.}],
\end{equation}
has given by Husimi~\cite{Husimi} in 1953 (and independently by
Kerner~\cite{Kerner} in 1958). Here, we give another method to deal
this problem by taking a time-dependent displacement transformation
\begin{equation}
D[f(t)]=\exp[f(t)a^{\dagger}-f^{*}(t)a],\label{eq:M}
\end{equation}
where function $f(t)$ is to be determined. It is ready to find the
relations
\begin{equation}
DaD^{\dagger}=a-f(t),\ {\rm and}\ Da^{\dagger}D^{\dagger}=a^{\dagger}-f^{*}(t).
\end{equation}
After the transformation, the effective Hamiltonian reads
\begin{eqnarray}
\tilde{H} & = & DHD^{\dagger}-iD\left(\frac{\partial}{\partial t}D^{\dagger}\right)\\
 & = & \omega a^{\dagger}a-(\omega f-F^{*}-i\dot{f})a^{\dagger}-(\omega f^{*}-F+i\dot{f}^{*})a\nonumber\\
 &   & +\omega|f|^{2}-(Ff+F^{*}f^{*})+\frac{i}{2}(f\dot{f}^{*}-\dot{f}f^{*}).
\end{eqnarray}
One finds that the Hamiltonian will be diagonalized if we choose suitable
function $f(t)$ satisfying\begin{subequations}
\begin{eqnarray}
\omega f-F^{*}-i\dot{f} & = & 0,\label{eq:f1}\\
\omega f^{*}-F+i\dot{f}^{*} & = & 0.\label{eq:f2}
\end{eqnarray}
\end{subequations}

Now we split $f(t)$ into real and imaginary parts
\begin{equation}
f(t)=-\left(\sqrt{\frac{m_{e}\omega}{2}}x_{c}+i\sqrt{\frac{1}{2m_{e}\omega}}p_{c}\right).
\end{equation}
Then Eqs.~(\ref{eq:f1}) and (\ref{eq:f2}) change into\begin{subequations}
\begin{eqnarray}
\dot{p}_{c} & = & -\left[m\omega^{2}x_{c}+\sqrt{\frac{m_{e}\omega}{2}}(F+F^{*})\right],\label{eq:f1-1}\\
\dot{x}_{c} & = & \frac{p_{c}}{m}+i\sqrt{\frac{1}{2m_{e}\omega}}(F-F^{*}).\label{eq:f2-1}
\end{eqnarray}
\end{subequations}
If $F$ is real, it is observable that $x_{c}$
and $p_{c}$ satisfy the classical Hamilton equation generating by
the classical Hamiltonian of a forced classical harmonic oscillator
\begin{equation}
\mathcal{H}_{c}=\frac{p_{c}^{2}}{2m_{e}}+\frac{1}{2}m_{e}\omega^{2}x_{c}^{2}+\tilde{F}(t)x_{c},
\end{equation}
with $\tilde{F}(t)=\sqrt{2m_{e}\omega}F(t)$.

In our case $F=F^{*}=\xi\cos\nu t$, then we obtain the solution
\begin{eqnarray}
\!\!\!\!\!\!\!\!\!\!\!x_{c}(t) \!\!& =\!\! & \!-\sqrt{\!\!\frac{2}{m_{e}\omega}}\!\frac{\omega\xi}{\omega^{2}\!-\!\nu^{2}}\!\cos\nu t\!+\!A\sin\omega t\!+\!B\cos\omega t \!,\\
\!\!\!\!\!\!\!\!\!\!\!p_{c}(t) \!\!& = \!\!& \!\!\sqrt{2m_{e}\omega}\!\frac{\nu\xi}{\omega^{2}\!-\!\nu^{2}}\!\sin\nu t\!+\!A\omega\!\cos\omega t\!-\!B\omega\!\sin\omega t \!,
\end{eqnarray}
where $A$ and $B$ are time-independent constants. For simplicity,
we take $A=B=0$ (corresponding to the special initial conditions
$x_{c}(0)=-\sqrt{2\omega/m_{e}}\xi/(\omega^{2}-\nu^{2})$ and $\dot{x}_{c}(0)=0$),
then the displacement operator reads
\begin{equation}
D=e^{-i[px_{c}(t)+xp_{c}(t)]},
\end{equation}
and the Hamiltonian changes into
\begin{equation}
H_{D}=\omega a^{\dagger}a+\mathcal{E}_{c},
\end{equation}
where
\begin{equation}
\mathcal{E}_{c}=-\frac{\omega\xi^{2}}{\omega^{2}-\nu^{2}}\cos^{2}\nu t.\label{eq:Hc}
\end{equation}
As shown in the following subsection, $\mathcal{E}_{c}$ corresponds
to the classical energy for the DHO.

\subsection{Equivalent quantum linear coordinate transformation}

In this subsection, it will shown that the time-dependent unitary
transformation in the former subsection corresponds to a quantum linear
coordinate transformation\cite{Rosen,Husimi}.

Now, we rewrite the Hamiltonian of the DHO as
\begin{eqnarray}
H & = & -\frac{\hbar^{2}}{2m_{e}}\nabla^{2}+\frac{1}{2}m_{e}\omega^{2}x^{2}+\tilde{F}(t)x.
\end{eqnarray}
Then we take a linear coordinate-translation transformation
\begin{equation}
x'=x-x_{c}(t),
\end{equation}
where the time-dependent c-number $x_{c}(t)$ satisfies the classical
Hamilton equation\begin{subequations}
\begin{eqnarray}
\dot{x}_{c} & = & \frac{p_{c}}{m_{e}},\\
\dot{p}_{c} & = & -m_{e}\omega^{2}x_{c}-\tilde{F}(t),
\end{eqnarray}
\end{subequations}i.e.,
\begin{equation}
m_{e}\ddot{x}_{c}+m_{e}\omega^{2}x_{c}+\tilde{F}(t)=0.\label{eq:Hamilton}
\end{equation}
Obviously, $x_{c}(t)$ describes the classical path of the DHO. It
is ready to obtain the following relations
\begin{equation}
x'=x-x_{c}(t),\ t'=t,\ \nabla'=\nabla,\ \frac{\partial}{\partial t'}=\frac{\partial}{\partial t}+\dot{x}_{c}\nabla'.
\end{equation}

As the consequence of the required covariance,
\begin{equation}
i\hbar\frac{\partial}{\partial t'}\psi'(x',t')=H'\psi'(x',t'),
\end{equation}
a transformation of the wave function is needed
\begin{equation}
\psi'(x',t')=\psi(x,t)e^{-i\phi},
\end{equation}
with
\begin{equation}
\phi=\frac{1}{\hbar}\left[p_{c}x'+\frac{1}{m_{e}}\int_{0}^{t'}p_{c}^{2}(\tau)d\tau\right].
\end{equation}
Here, the Hamiltonian in the new reference frame reads
\begin{eqnarray}
H' & = & -\frac{\hbar^{2}}{2m}\nabla'^{2}+\frac{1}{2}m\omega^{2}x'^{2}+\mathcal{H}_{c}(t),
\end{eqnarray}
where
\begin{eqnarray}
\mathcal{H}_{c}(t) & = & \frac{p_{c}^{2}}{2m_{e}}+\frac{1}{2}m_{e}\omega^{2}x_{c}^{2}+\tilde{F}(t')x_{c},\label{eq:classical_H}
\end{eqnarray}
is the classical Hamiltonian of the DHO. It is found that the DHO
moves as a free harmonic oscillator in this new reference frame.

For our case $F=\hbar\xi\cos\nu t$, we also take the solution of the Hamilton
equation~(\ref{eq:Hamilton}) as
\begin{equation}
x_{c}(t)=-\sqrt{\frac{2\omega}{\hbar m_{e}}}\frac{\xi\cos\nu t'}{\omega^{2}-\nu^{2}}.
\end{equation}
Following Eq.~(\ref{eq:classical_H}), we obtain the classical energy
of the DHO
\begin{equation}
\mathcal{E}_{c}(t')=-\frac{\omega\xi^{2}}{\omega^{2}-\nu^{2}}\cos^{2}\nu t',
\end{equation}
which is the exact time-dependent function defined in Eq.~(\ref{eq:Hc}). Here,
we have neglected a time-independent constant $\omega\xi^{2}\nu^{2}/(\omega^{2}-\nu^{2})^{2}$
and taken $\hbar=1$. Consequently, this quantum linear coordinate
transformation is equivalent with the time-dependent displacement
$D(t)$ defined in Eq.~(\ref{eq:M}).

\section{effective spin precession in two-dimensional quantum dot system}

\subsection{Diagonalization of the vibration potential}

By defining creation and annihilation operators
\begin{equation}
x=\frac{1}{\sqrt{2m_{e}\omega_{x}}}(a^{\dagger}+a),\ p_{x}=i\sqrt{\frac{m_{e}\omega_{x}}{2}}(a^{\dagger}-a),
\end{equation}
\begin{equation}
y=\frac{1}{\sqrt{2m_{e}\omega_{y}}}(b^{\dagger}+b),\ p_{y}=i\sqrt{\frac{m_{e}\omega_{y}}{2}}(b^{\dagger}-b),
\end{equation}
we rewrite the vibration Hamiltonian (\ref{Hv}) of the electron in 2D QD system as
\begin{eqnarray*}
H_{v} & = & \omega_{x}a^{\dagger}a+\omega_{y}b^{\dagger}b-\chi(a^{\dagger}b+ab^{\dagger})
\end{eqnarray*}
where
\begin{eqnarray}
\omega_{x}^{2} & = & \tilde{\omega}_{x}^{2}+\omega_{c}^{2}\sin^{2}\varphi_{B},\\
\omega_{y}^{2} & = & \tilde{\omega}_{y}^{2}+\omega_{c}^{2}\cos^{2}\varphi_{B},\\
\omega_{c} & = & \frac{eB}{m_{e}c}.
\end{eqnarray}
and we have taken the rotating wave approximation since
\begin{equation}
\chi=\omega_{c}\sin2\varphi_{B}\sqrt{\frac{\omega_{c}^{2}}{\omega_{x}\omega_{y}}}\ll\omega_{x(y)}.
\end{equation}
And the SOC Hamiltonian~$H_{{\rm so}}$ and the driven part~$V(t)$ change into
\begin{eqnarray}
H_{{\rm so}}& = & i\sqrt{\frac{m_{e}\omega_{y}}{2}}(\alpha_{R}\sigma_{x}+\alpha_{D}\sigma_{y})(b^{\dagger}-b)\nonumber \\
&  & -i\sqrt{\frac{m_{e}\omega_{x}}{2}}(\alpha_{R}\sigma_{y}+\alpha_{D}\sigma_{x})(a^{\dagger}-a),
\end{eqnarray}
and\begin{widetext}
\begin{equation}
V(t)=eE\left[\sqrt{\frac{1}{2m_{e}\omega_{x}}}(a^{\dagger}+a)\cos\varphi_{E}+\sqrt{\frac{1}{2m_{e}\omega_{y}}}(b^{\dagger}+b)\sin\varphi_{E}\right]\cos\nu t,
\end{equation}
\end{widetext}
respectively. Hereafter, we let $\hbar=1$ for simplicity.

It is convenient to diagonalize $H_{v}$ by defining two new modes,
\begin{equation}
\left(\begin{array}{c}
A\\
B
\end{array}\right)=\left(\begin{array}{cc}
\cos\frac{\vartheta}{2} & -\sin\frac{\vartheta}{2}\\
\sin\frac{\vartheta}{2} & \cos\frac{\vartheta}{2}
\end{array}\right)\left(\begin{array}{c}
a\\
b
\end{array}\right),
\end{equation}
where
\begin{equation}
\cos\vartheta=\frac{\Delta/2}{\sqrt{\Delta^{2}/4+\chi^{2}}},\ \sin\vartheta=\frac{\chi}{\sqrt{\Delta^{2}/4+\chi^{2}}},
\end{equation}
and~$\Delta=\omega_{x}-\omega_{y}$.
Then the total Hamiltonian of the 2D QD system $H=H_{v}+H_{s}+H_{{\rm so}}+V(t)$ changes
into
\begin{equation}
H_{v}=\omega_{A}A^{\dagger}A+\omega_{B}B^{\dagger}B,
\end{equation}
\begin{equation}
H_{s}=\frac{1}{2}g\mu_{B}\vec{B}\cdot\vec{\sigma},
\end{equation}
\begin{eqnarray}
\!\!\!\!\!\!V(t) & = &\! \xi_{A}(A^{\dagger}\!+\!A)\cos\nu t\!+\!\xi_{B}(B^{\dagger}+B)\cos\nu t,
\end{eqnarray}
and
\begin{widetext}
\begin{eqnarray}
H_{{\rm so}} & = & i\left[\sqrt{\frac{m_{e}\omega_{y}}{2}}(\alpha_{R}\sigma_{x}+\alpha_{D}\sigma_{y})\cos\frac{\vartheta}{2}-\sqrt{\frac{m_{e}\omega_{x}}{2}}(\alpha_{R}\sigma_{y}+\alpha_{D}\sigma_{x})\sin\frac{\vartheta}{2}\right](B^{\dagger}-B)\nonumber \\
 &  & -i\left[\sqrt{\frac{m_{e}\omega_{y}}{2}}(\alpha_{R}\sigma_{x}+\alpha_{D}\sigma_{y})\sin\frac{\vartheta}{2}+\sqrt{\frac{m_{e}\omega_{x}}{2}}(\alpha_{R}\sigma_{y}+\alpha_{D}\sigma_{x})\cos\frac{\vartheta}{2}\right](A^{\dagger}-A).
\end{eqnarray}
\end{widetext}
Here, the frequencies of the new two modes are
\begin{eqnarray}
\omega_{A} & = & \frac{\omega_{x}+\omega_{y}}{2}+\sqrt{\frac{\Delta^{2}}{4}+\chi^{2}},\\
\omega_{B} & = & \frac{\omega_{x}+\omega_{y}}{2}-\sqrt{\frac{\Delta^{2}}{4}+\chi^{2}},
\end{eqnarray}
and corresponding driving strength of the electric field,
\begin{eqnarray}
\!\!\!\!\!\!\!\!\!\!\xi_{A}\!\! & = &\!\!\frac{eE}{\sqrt{2m_{e}}}\![\frac{1}{\sqrt{\omega_{x}}}\!\cos\frac{\vartheta}{2}\!\cos\varphi_{E}\!-\!\!\frac{1}{\sqrt{\omega_{y}}}\sin\frac{\vartheta}{2}\!\sin\varphi_{E}],\\
\!\!\!\!\!\!\!\!\!\!\xi_{B}\!\! & = &\!\!\frac{eE}{\sqrt{2m_{e}}}\![\frac{1}{\sqrt{\omega_{x}}}\!\sin\frac{\vartheta}{2}\!\cos\varphi_{E}\!+\!\frac{1}{\sqrt{\omega_{y}}}\cos\frac{\vartheta}{2}\!\sin\varphi_{E}].
\end{eqnarray}

\subsection{Effective spin-controlling Hamiltonian}

Just like the one-dimensional case, we take a similar unitary transformation
\begin{eqnarray}
D[t] & = & \exp\left[f_{A}(t)A^{\dagger}+f_{B}(t)B^{\dagger})-{\rm h.c.}\right].
\end{eqnarray}
where
\begin{eqnarray}
f_{A} & = & \frac{\xi_{A}}{\omega_{A}^{2}-\nu^{2}}(\omega\cos\nu t-i\nu\sin\nu t),\\
f_{B} & = & \frac{\xi_{B}}{\omega_{B}^{2}-\nu^{2}}(\omega\cos\nu t-i\nu\sin\nu t).
\end{eqnarray}
Then the Hamiltonian changes into $H=H_{v}+H_{s}+H_{{\rm so}}+H_{{\rm flip}}$,
where
\begin{equation}
H_{v}=\omega_{A}A^{\dagger}A+\omega_{B}B^{\dagger}B,
\end{equation}
and
\begin{eqnarray}
\!\!\!\!\!\!\!\!\!\!\!\!\!\!H_{{\rm flip}}\!\!\! & = & \!\!\![G_{1}(\alpha_{R}\sigma_{x}\!+\!\alpha_{D}\sigma_{y})\!-\!G_{2}(\alpha_{R}\sigma_{y}\!+\!\alpha_{D}\sigma_{x})]\!\sin\nu t\!,
\end{eqnarray}
with
\begin{widetext}
\begin{eqnarray}
G_{1} & = & \frac{1}{2}\eta_{B}(\sqrt{\frac{\omega_{y}}{\omega_{x}}}\sin\frac{\vartheta}{2}\cos\frac{\vartheta}{2}\cos\varphi_{E}+\cos^{2}\frac{\vartheta}{2}\sin\varphi_{E})-\frac{1}{2}\eta_{A}(\sqrt{\frac{\omega_{y}}{\omega_{x}}}\sin\frac{\vartheta}{2}\cos\frac{\vartheta}{2}\cos\varphi_{E}+-\sin^{2}\frac{\vartheta}{2}\sin\varphi_{E}),\\
G_{2} & = & \frac{1}{2}\eta_{B}\left(\sqrt{\frac{\omega_{x}}{\omega_{y}}}\sin\frac{\vartheta}{2}\cos\frac{\vartheta}{2}\sin\varphi_{E}+\sin^{2}\frac{\vartheta}{2}\cos\varphi_{E}\right)-\frac{1}{2}\eta_{A}\left(\sqrt{\frac{\omega_{x}}{\omega_{y}}}\sin\frac{\vartheta}{2}\cos\frac{\vartheta}{2}\sin\varphi_{E}-\cos^{2}\frac{\vartheta}{2}\cos\varphi_{E}\right).
\end{eqnarray}
\end{widetext}
Here, the additional spin-flipping term $H_{{\rm flip}}$ is generated
by the electric field mediated by SOC and
\begin{equation}
\eta_{A(B)}=\frac{2\nu eE}{\omega_{A(B)}^{2}-\nu^{2}}.
\end{equation}

For most case, the azimuthal dependence of $H_{{\rm flip}}$ about
$\varphi_{E}$ is complicate. In experiment, the external static magnetic
field is week $\chi\ll\omega$, and then the influence of the vector
potential $\vec{A}(\vec{r})$ can be neglected, i.e., $\vartheta\rightarrow0$
and $\omega_{A(B)}\approx \tilde{\omega}_{x(y)}$. If the 2D harmonic well
is isotropic $\tilde{\omega}_{x}=\tilde{\omega}_{y}=\omega$, $H_{{\rm flip}}$
will get highly simplified as
\[
H_{{\rm flip}}=\frac{1}{2}g\mu_{B}\left[\vec{B}_{R}(t)+\vec{B}_{D}(t)\right]\cdot\vec{\sigma},
\]
where $\vec{B}_{R}(t)=B_{R}\sin\nu t(\sin\varphi_{E},-\cos\varphi_{E},0)$
is magnetic fields induced by Rashba spin-orbit coupling and the magnetic
field induced by Dresselhaus coupling is $\vec{B}_{D}(t)=B_{D}\sin\nu t(-\cos\varphi_{E},\sin\varphi_{E},0)$
with strengths
\begin{equation}
B_{R(D)}=\frac{2\nu eE\alpha_{R(D)}}{g\mu_{B}\left(\omega^{2}-\nu^{2}\right)}.
\end{equation}

\end{document}